\documentclass[epj,twocolumn]{webofc}
\usepackage[varg]{txfonts}   
\usepackage{graphicx}

\newcommand{\muHz}{\;{\rm {\mu}Hz}}

\newcommand{\dscu}{\delta\;{\rm Scuti}}
\newcommand{\gdor}{\gamma\;{\rm Doradus}}
\newcommand{\numax}{\nu_{\rm max}}
\newcommand{\dnu}{\Delta \nu}
\newcommand{\fmin}{f_{\rm min}}
\newcommand{\fmax}{f_{\rm max}}
\newcommand{\fminfmax}{f_{\rm min}\_f_{\rm max}}
\newcommand{\amax}{a_{\rm max}}

\wocname{epj}
\woctitle{Seismology of the Sun and the Distant Stars 2016}
\begin{document}
\title{What CoRoT tells us about $\delta$ Scuti stars}
%
\subtitle{existence of a regular pattern and seismic indices to characterize stars}

\author{\firstname{Eric} \lastname{Michel}\inst{1}\fnsep\thanks{\email{Eric.Michel at obspm.fr}} 
\and
        \firstname{Marc-Antoine} \lastname{Dupret}\inst{2}
\and
        \firstname{Daniel} \lastname{Reese}\inst{1}
\and
        \firstname{Rhita-Maria} \lastname{Ouazzani}\inst{3}
\and
        \firstname{Jonas} \lastname{Debosscher}\inst{4}
\and
        \firstname{Antonio} \lastname{Garc\'ia Hern\'andez}\inst{5,6}
\and
        \firstname{Kevin} \lastname{Belkacem}\inst{1}
\and
        \firstname{Reza} \lastname{Samadi}\inst{1}
\and
        \firstname{S\'ebastien} \lastname{Salmon}\inst{2}
\and
        \firstname{Juan Carlos} \lastname{Suarez}\inst{6}
\and
        \firstname{Sebastia} \lastname{Barcel\'o Forteza}\inst{7,8}
}

\institute{Observatoire de Paris, LESIA, UMR8109, Universit\'e Pierre et Marie Curie, Universit\'e
Paris Diderot, PSL 
\and
           Institut d'Astrophysique et Géophysique, University of Liège, Belgium 
\and
           Stellar Astrophysics Centre, Department of Physics and Astronomy, Aarhus University, Ny Munkegade 120, DK-8000 Aarhus C, Denmark
\and
           Instituut voor Sterrenkunde, KU Leuven, Celestijnenlaan 200B, 3001, Leuven, Belgium
\and
            Instituto de Astrofísica e Ciências do Espaco, Universidade do Porto, CAUP, 
Rua das Estrelas, PT4150-762 Porto, Portugal
\and
          Dept. F\'isica Te\'orica y del Cosmos. University of Granada. 18071. Granada, Spain 
\and
           Instituto de Astrof\'isica de Canarias, 38200 La Laguna, Tenerife, Spain 
\and
           Departamento de Astrof\'isica, Universidad de La Laguna, 38206 La
Laguna, Tenerife, Spain 
          }
\abstract{%
 Inspired by the so appealing example of red giants, where going from a handful 
of stars to thousands revealed the structure of the eigenspectrum, we inspected a 
large homogeneous set of around 1860 $\dscu$  stars observed with CoRoT. This unique data 
set 
reveals a common regular pattern which appears to be in agreement with island modes 
featured by theoretical non-perturbative treatments of fast rotation. 
The comparison of these data with models and linear stability calculations  
suggests that spectra can be fruitfully characterized to first 
order by a few parameters which might play the role 
of seismic indices for $\dscu$ stars, as $\dnu$ and $\numax$ do for red giants.
The existence of this pattern offers an observational support for guiding further theoretical works 
on fast rotation. 
It also provides a framework for further investigation of the observational material 
collected by CoRoT (\cite{Baglin2006}) and {\it Kepler} (\cite{Borucki2010}). 
Finally, it sketches out the perspective of using $\dscu$ 
 stars pulsations for ensemble asteroseismology.  
}
\maketitle
%
\section{Introduction}
\label{intro}
$\dscu$ stars constitute a large class of pulsating stars representative 
of chemically normal intermediate mass stars on and near the main sequence 
(see eg \cite{Breger2000}, \cite{Rodriguez2001}). 
Their seismic exploitation however meets major difficulties often refered to as
'the mode identification problem', 'the fast rotation treatment', and 'the selection effects'. 
These various expressions refer to two main difficulties.  
First, $\dscu$ stars, as chemically normal A and early F stars are characterized by a 
large rotation rate (\cite{Royer2014}) and the theoretical modelling of their pulsation spectra cannot rely
on classical perturbative approaches (\cite{Lignieres2006},\cite{Reese2006}).
Then, although we understand the process responsible for
their pulsational instability, we have very little insight about the process responsible for the amplitude 
limitation (see however recent studies by \cite{Barcelo2015} and \cite{Bowman2016} )and thus no 
clue about how amplitudes are distributed between modes and for
different stars.

However, we do know a couple of things about these stars. 
First, there is increasing evidence that periodicities or regular spacings
can be found in $\dscu$ spectra (\cite{GarciaH2013}, \cite{Paparo2016}, \cite{Breger2011})
and
recently, \cite{GarciaH2015} demonstrated that this spacing is a good proxy of the mean density
just as the large separation $\dnu$ is for solar-like pulsators.

Then, linear stability calculations provide reliable results (\cite{Dupret2004}) and it is appealing
to use them to characterize stars in terms of mass range or evolution stage  
(see e.g. \cite{Michel1999} for stars in clusters and \cite{Zwintz2014} in the case 
of pre-main sequence $\dscu$ stars). 

In the present paper, we use a large set of homogeneous spectra observed with CoRoT (\cite{Baglin2016})
to revisit these questions and
see what CoRoT data tell us about $\dscu$ stars.

\section{The observational sample and the determination of $\fmin$, $\fmax$ and $\amax$.}
\label{sec-1}

The automated supervised classification of variable stars in the CoRoT programme 
(ASCVC hereafter, \cite{Debosscher2009}) brings about 1860 objects classified as $\dscu$ stars with a
probability higher than 80$\%$. In comparison, catalogues before the space-photometry era gathered about
700 objects (\cite{Rodriguez2000}), among which a large fraction had been discovered by 
large surveys like the Hipparcos 
(\cite{Perryman1997}), OGLE (\cite{Udalski1997}) and MACHO (\cite{Alcock2000}) projects. 
The present CoRoT sample is thus very valuable in terms of number of objects and also in terms of 
homogeneity.

We computed the Fourier spectrum for each of these light-curves and we set to
zero-amplitude the parts of the spectra possibly hampered by intrumental/environmental artefacts induced by
the orbital period (see \cite{Auvergne2009}). It consists in narrow intervals around the 
frequency of the orbital period plus 
its harmonics, each of them associated with a few daily aliases as it can be seen in figure~\ref{fig-6}.
For each spectrum, we also exclude from our study the part below $f_{Lcut}=25\muHz$ (also set to zero-amplitude)
in order to avoid the influence on our analysis of possible power of instrumental or environmental 
origin at low frequency. 

We define a limit amplitude criterion for peaks to be considered. Here we take the maximum between 
10 times the mean amplitude level and the
amplitude of the highest peak divided by 8. This last constraint aims at avoiding artefacts of large
amplitude peaks convolved by the observational window. 

Then, we determine for each spectrum the range of detected signal, noted $[\fmin,\fmax]$, as 
the frequency range encompassing all peaks satisfying the previous amplitude criterion.

We also use these spectra to produce an index characterizing the amplitude of the oscillations. We
considered two versions of this index. One is simply the amplitude of the highest peak. The second
is the square root of the quadratic sum  of amplitudes of all peaks satisfying our amplitude criterion.
Interestingly, the results were found to vary at most by a factor two from one version to the other. 
In the present work, all results
for $\amax$ refer to the square root of the quadratic sum, which we expect to be more stable
a measurement and more representative of the energy involved in pulsation. 

\begin{figure}[h]
\centering
\includegraphics[width=\hsize,clip]{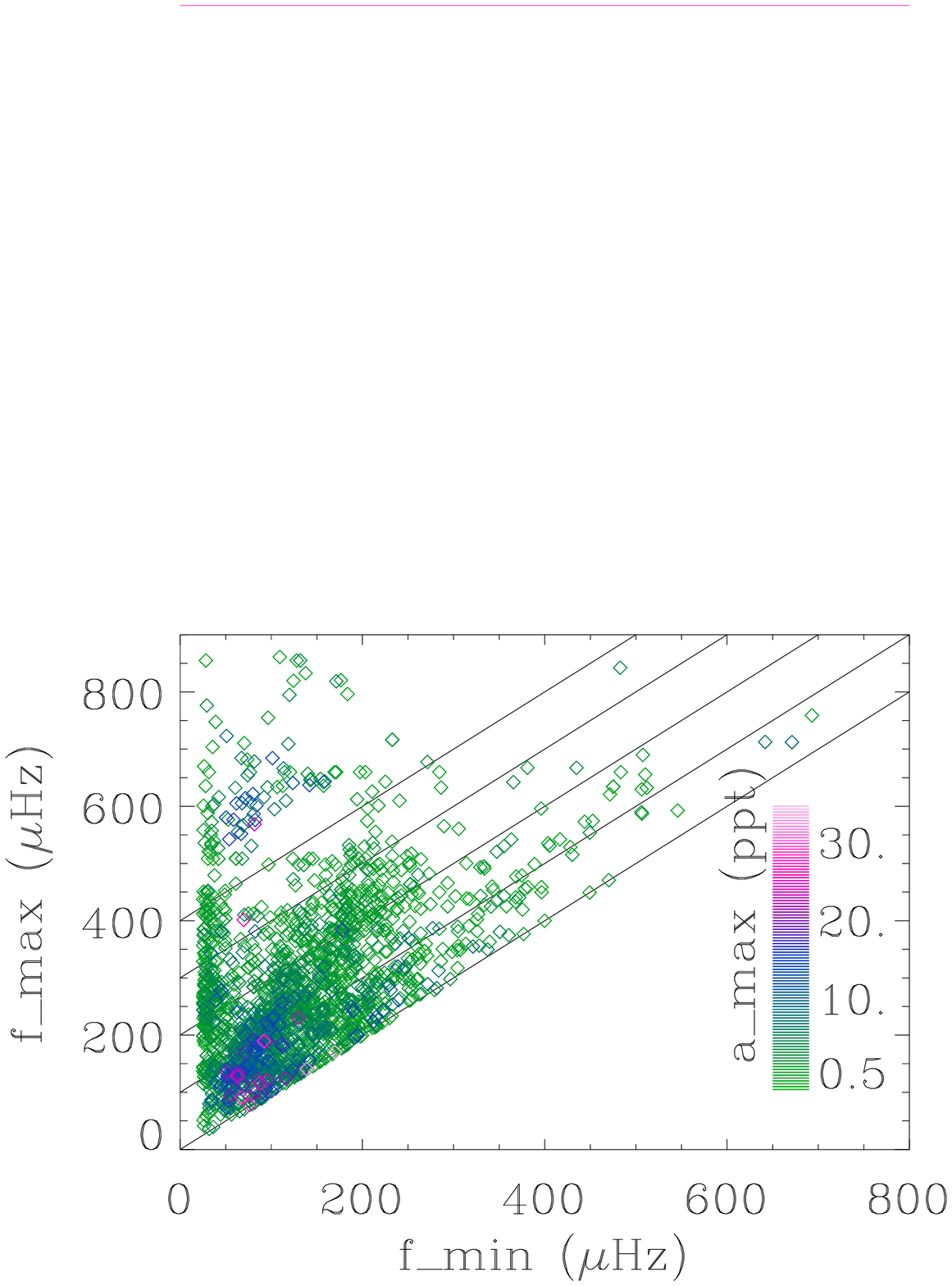}
\caption{Diagram $\fminfmax$ for the set of CoRoT stars described in the text. Oblique lines
indicate equal frequency width $\fmax-\fmin$. The values mentioned on the y-axis indicate both
the $\fmax$ value for a horizontal line and the $\fmax-\fmin$ width for an oblique line.
Values of $\amax$ are given by the colour code}
\label{fig-1}       
\end{figure}

The results are presented in figure~\ref{fig-1} and their interpretation in the light of theoretical
models is discussed in Sect.~\ref{sec-3}.
 
\section{Theoretical estimates of $\fmin$ and $\fmax$}
\label{sec-2}

\begin{figure}[h]
\centering
\includegraphics[width=\hsize,clip]{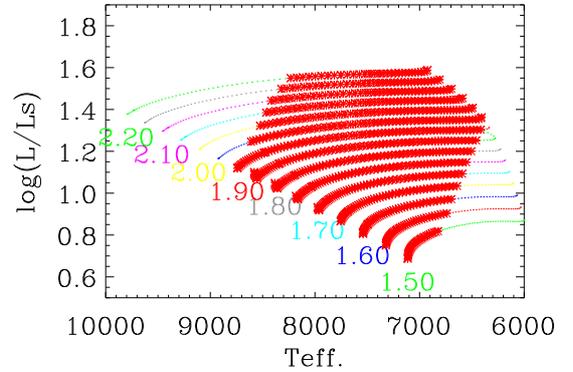}
\caption{Hertzsprung-Russell diagram featuring models described in the text. Models showing at least
one unstable mode are marked by red star symbols delimiting the theoretical instability strip. For each
sequence, mass is indicated in solar units.}
\label{fig-2}       
\end{figure}

We used a grid of theoretical models representative of the whole $\dscu$ stars instability strip for
the main sequence evolution stage (see figure~\ref{fig-2}).
This grid is the one used and described in \cite{Dupret2004}. 
The models have been computed with the code CLES (\cite{Scuflaire2008}) 
The physics of the models is standard and the 
only specific aspects of interest at the level of the present study are: 
the use of overshooting  with $\alpha_{ov}=0.2 H_p$ and the mixing length 
parameter which has been set to the solar calibrated value 1.8.
The metallicity (Z=0.02) had been chosen as representative of the solar one in \cite{Dupret2004}. 
Here again, the change for a more up-to-date value would not change significantly our results. 
The linear stability of the modes has been obtained following (\cite{Dupret2004},
\cite{Grigahcene2005}).

\begin{figure}[h]
\centering
\includegraphics[width=\hsize,clip]{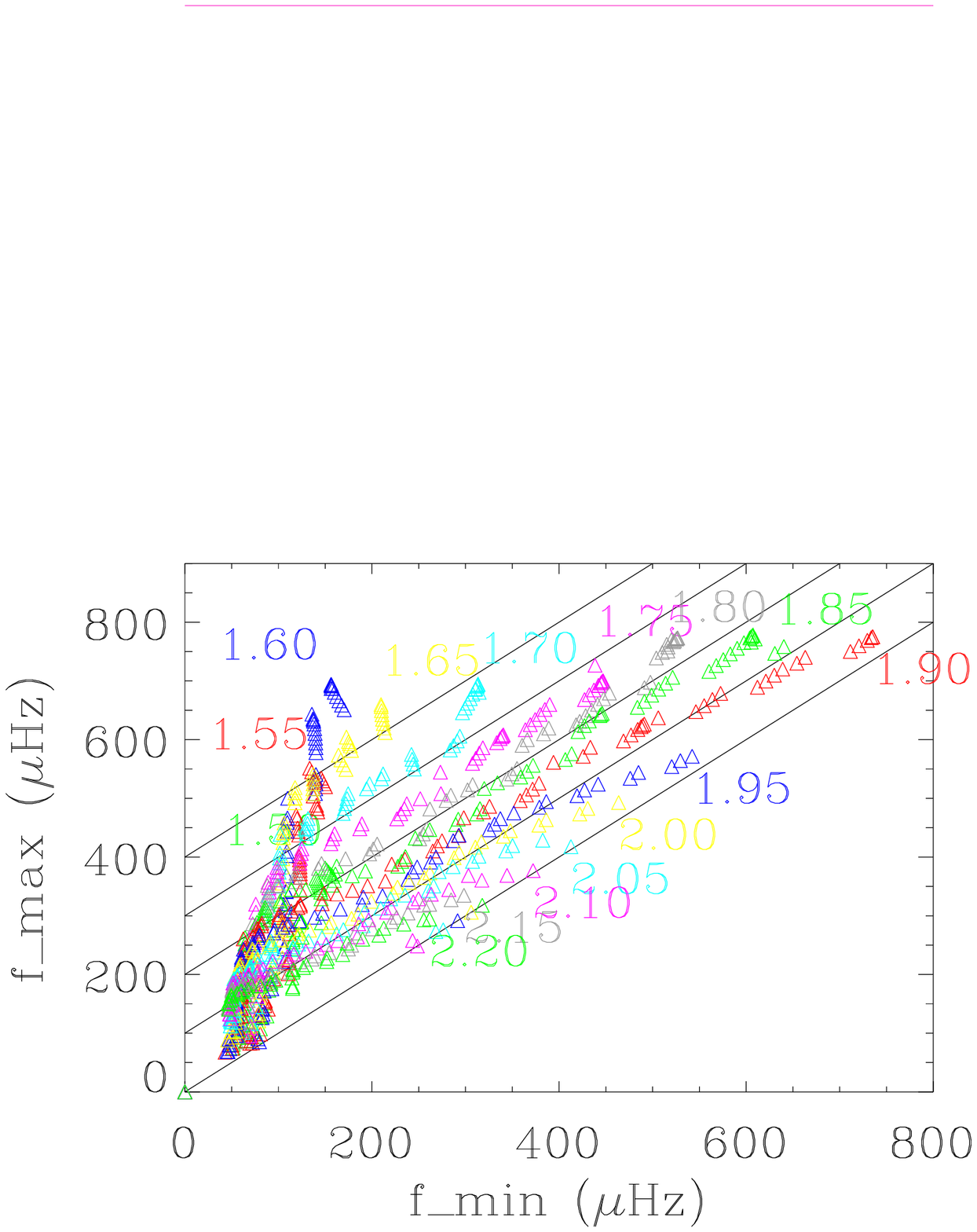}
\caption{Diagram $\fminfmax$, as in figure~\ref{fig-1}, but for sequences of models illustrated in
figure~\ref{fig-2} and described in the text. The colours stand for sequences of models of a given mass.}
\label{fig-3}       
\end{figure}

These linear stability calculations are used to derive theoretical 
counterparts of the $\fmin$ and $\fmax$ values obtained for 
observed stars in Sect.~\ref{sec-1}. 
They have been determined on modes of degree $l=0,1,2$ and are illustrated in figure~\ref{fig-3}. 

When looking at figure~\ref{fig-3}, it is clear that, for sequence of models of given mass,
the position in the 
$\fminfmax$ diagram changes under two effects. One is the decrease of mean density 
(or the decrease of $\Delta {\nu}_0 \propto (G M/R^3)^{1/2}$) with evolution. The second is associated with the 
gradual decrease of the radial orders of unstable modes, as the star crosses the instability strip from 
the blue to the red border, as described in Ref Dupret 2004.
Both effects globally induce a decrease of $\fmin$ and $\fmax$  with age on the main sequence.

It is also worth noticing that, evolution sequences entering the instability strip through the 
so-called blue border (i.e. masses higher than $1.9M_\odot$) correspond, in the $\fminfmax$ 
diagram, to sequences starting on the
$(\fmin=\fmax)$ central axis, i.e. with a very narrow frequency range of unstable modes.

On the contrary, evolution sequences starting on the ZAMS
in the instability strip (i.e. masses between 1.5 and 1.9$M_\odot$) correspond,
in the $\fminfmax$ diagram, to sequences starting at high $\fmax - \fmin$ values,
i.e. with a larger frequency range of unstable modes and thus farther from the $(\fmin=\fmax)$ central axis.

Finally, we can also notice that, with the evolution, all sequences tend to converge at low 
$\fmin$ and $\fmax$ in the diagram, those going out of the instability strip
through the red border corresponding to sequences terminating on the $(\fmin=\fmax)$ central axis.

We take as a work hypothesis that the rotation will not impact drastically these 
theoretical estimates of $\fmin$ and $\fmax$ values. Rotational splitting is expected to extend
the observed range of observed peaks by an amount which is difficult to estimate precisely in the 
present state of our knowledge, but should
be, in the worst case, of the order of a few times the rotation frequency, 
i.e. a few times 10 to 20$\muHz$ for the fastest rotating objects.
This is not negligible, but if this extension remains below 50$\muHz$ for most of the stars, 
the $\fminfmax$ diagram remains discriminent in terms of evolution stage and mass range, as
can be seen in figure~\ref{fig-3}.

\section{Comparison of observational and theoretical $\fminfmax$ diagrams}
\label{sec-3}

When comparing figure~\ref{fig-1} and figure~\ref{fig-3} in the light of previous remarks, 
we notice a few differences and similarities. The set of observed stars shows 
components in the $\fminfmax$ diagram which are
not present for the models. The main difference consists in a vertival ridge 
observed at low $\fmin$ (below 50$\muHz$) in figure~\ref{fig-1}. 
This accumulation ridge could be due to an edge effect induced 
by the processing of spectra described in Sect.~\ref{sec-1}, when we fix the $f_{Lcut}$ value  
to prevent our analysis from the influence of possible low frequency noise. 
However, we should keep in mind that this 
ridge is also the place where we should expect hybrid $\dscu$-$\gdor$ stars. 
In addition to the modes characterizing $\dscu$ stars,
these objects present low-frequency pulsation
modes not considered unstable in our models. 
A closer inspection of these spectra will be necessary to explore this possibility. 

The second component presented by the observed set of stars and not by the models is much less important
in numbers. It appears as a cloud of points around the upper left corner of figure~\ref{fig-1}, i.e. stars
with $\fmin$ values bellow 200$\muHz$ and $\fmax$ values between 500 and 900$\muHz$. 
As it reads from the oblique lines in figure~\ref{fig-1}, 
this corresponds to extremely large estimates of frequency range value 
($\fmax-\fmin > 500 \muHz$). The inspection of several spectra suggests that, to a large extent,
these values could be due to the poor handling of particularly severe window-artefacts by the data processing 
presented in Sect.~\ref{sec-1}. This is supported by the presence of a significant fraction of
spectra with high $\amax$ values among those points. 

Beside this, the $\fminfmax$ diagram for observed stars and the one for models look in 
reasonably good agreement, with a scarce distribution of points at high $\fmin$ and $\fmax$
values, where evolution sequences are highly spread in mass and age in figure~\ref{fig-3}, and 
a denser concentration of points at low values, where the evolution sequences tend to converge.

\subsection{ Amplitude of the oscillations versus evolution.}
\label{sec-30}
It is worth noticing that the distribution of $\amax$ shows a clear gradient (from 
0.1 to 40 part-per-thousand, ppt hereafter) with increasing amplitude values 
toward low values of $\fmin$ and $\fmax$. 
As we already commented, according to the theoretical diagram,
 this domain of the 
$\fminfmax$ diagram is expected to host rather evolved stars.

The fact that evolved $\dscu$ stars tend to show higher amplitudes than main sequence ones has been 
observed for long (see e.g. \cite{Rodriguez2001}),
but to our knowledge, this trend has never been observed 
with such a resolution in amplitude and on such a large sample of objects. 
In the case of {\it Kepler} data, \cite{Balona2011} considered about 1570 objects, but a large fraction of
them ($\sim$1150) were observed in the so-called long cadence mode, i.e. with a 30 minutes sampling time 
which is not suited to address frequencies higher than 270$\muHz$. 
These data will thus be very helpful to study $\dscu$ stars, but mostly evolved ones.

\subsection{ Evidence of a regular frequency pattern in early main sequence $\dscu$ stars.}
\label{sec-31}
Models off the main sequence are not produced here, but we have inspected several such sequences,
and they are always found in the domain of low $\fmin$ and $\fmax$. 
Even with $\alpha_{ov}=0$ (no overshooting), which corresponds to the shortest extension of
the main sequence, post-main sequence models seem to remain
below the $\fmax=400\muHz$ limit. 

\begin{figure}[h]
\centering
\includegraphics[width=\hsize,clip]{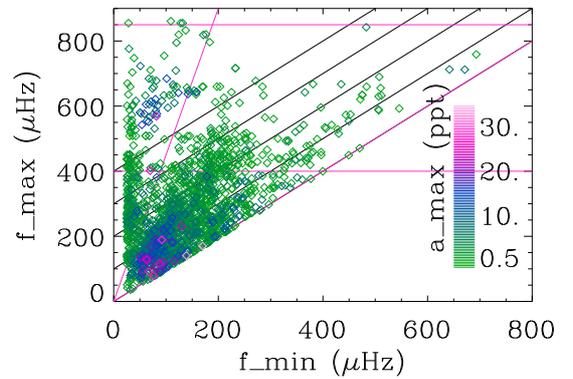}
\caption{Same as in figure~\ref{fig-1} but with purple lines marking the selection of the set of early
main sequence objects with $\fmax > 400\muHz$ and $\fmax/\fmin < 4.5$, as discussed in Sect.~\ref{sec-31}. 
} 
\label{fig-4}       
\end{figure}

\begin{figure}[h]
\centering
\includegraphics[width=\hsize,clip]{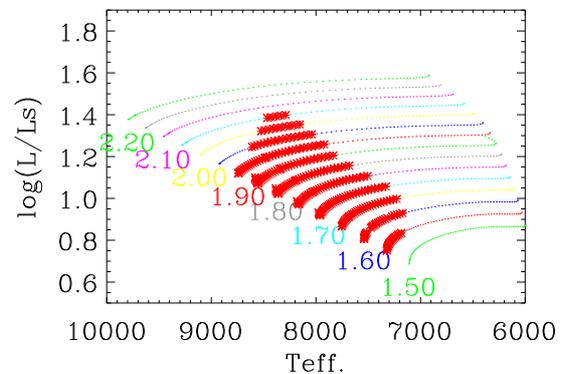}
\caption{Same as in figure~\ref{fig-2} but here only unstable models from the selection 
described in Sect.~\ref{sec-31} and illustrated in figure~\ref{fig-4} are marked by red symbols. 
} 
\label{fig-5}       
\end{figure}

We now consider stars with $\fmax > 400 \muHz$, i.e. for which, according to models,
$\fmin$ and $\fmax$ values correspond unambiguously 
to main sequence models, whatever the amount of overshooting considered. 
As illustrated in figure~\ref{fig-4} , we also avoid spectra with dubious values as 
discussed in Sect.~\ref{sec-1}. 
The domain of the HR diagram corresponding to this selection in the case
of models presented here is illustrated in figure~\ref{fig-5}.

\begin{figure*}
\centering
\includegraphics[width=\hsize,clip]{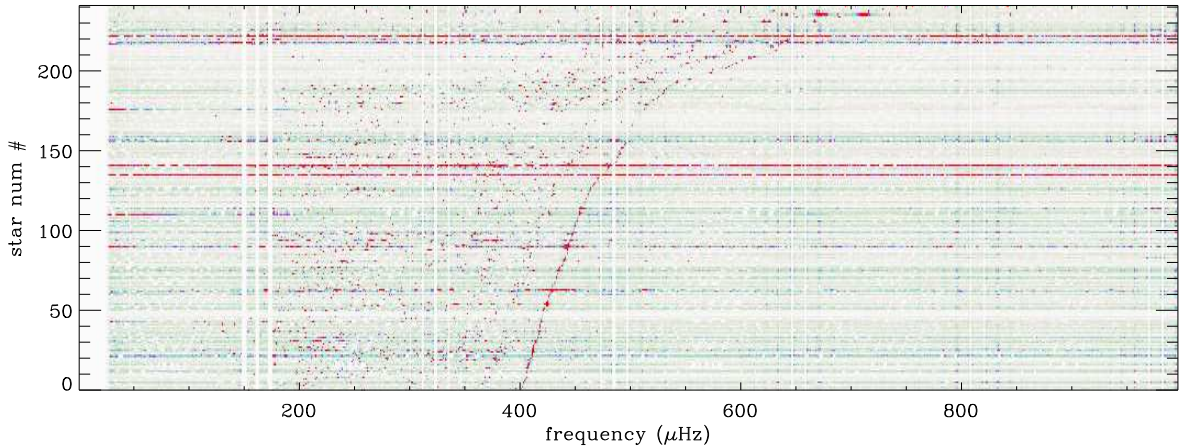}
\caption{
Each horizontal line of this image is an grey scale coded amplitude spectrum of one
of the stars belonging to the set of early main sequence stars selected as described
in Sect.~\ref{sec-31} and illustrated in figure~\ref{fig-4}. Each spectrum is truncated in amplitude
to values lower than a common
limit value (here $2. 10^{-4}$). The spectra clearly show thin vertical dark ridges resulting
from  setting to zero-amplitudes
parts of the spectra associated with the orbital period and its harmonics, plus day-aliases around them
as explained in Sect.~\ref{sec-1}.
}
\label{fig-6}       
\end{figure*}

For this set of about 200 stars, we build an image presented in figure~\ref{fig-6}, where
each amplitude spectrum is a line where amplitude is limited to an arbitrary value $a_{thr}$ 
(here $a_{thr}=0.2 ppt$) 
and coded in grey scale. The full image thus presents more than 200 spectra, sorted
by increasing $\fmax$ value from the bottom to the top.

In figure~\ref{fig-6}, we distinguish a few ridges approximately parallel to the ridge drawn by 
peaks associated with $\fmax$ values. 
These ridges separated by a few tens of $\muHz$ 
immediately recall the quasiregular pattern of axisymmetric island modes as described by nonperturbative
calculation of fast rotators (\cite{Reese2008} and \cite{Ouazzani2015}). 
They also recall the various detection of spacings of the order of the large separation in
$\dscu$ stars (\cite{GarciaH2015},\cite{Paparo2016})

The pattern associated with these ridges has to be common to a sufficient fraction of our 
stellar sample to show up in figure~\ref{fig-6}. This suggests 
that this quasiregular pattern of peaks has to be, to some extent, independent of rotation, 
which necessarily varies from one star to another.

Axisymmetric modes are good candidates to explain this pattern as we will show hereafter.

An appealing idea is that the eigenspectra of axisymmetric modes of the different stars could be, 
at first order, homologous
and thus distributed according to a common pattern of peaks, just multiplied by a different value
of the large separation (or mean density).   

\begin{figure*}
\centering
\includegraphics[width=\hsize,clip]{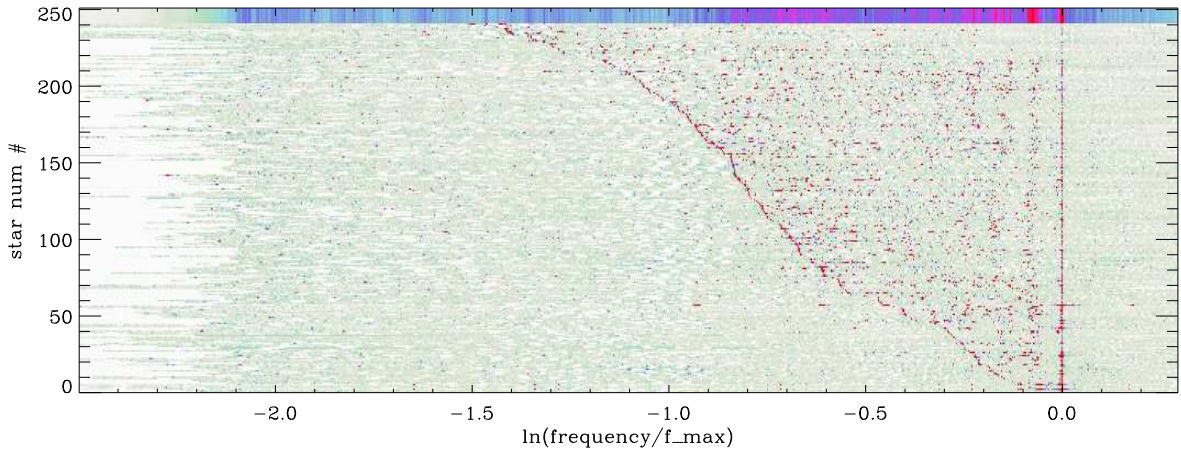}
\caption{
Same as in figure~\ref{fig-6} but abscissae of the spectra have been normalized by $\fmax$ and
converted in logarithm as described in the text. In addition, spectra have been ordered by increasing
$\fmax/\fmin$ value from the bottom to the top. The upper part of the image has been extended to host
a few additional lines showing (also in grey scale) the mean of all individual spectra.
}
\label{fig-7}       
\end{figure*}

In order to test a bit further this idea, we rescaled the (observational) spectra, taking in abscissae 
the logarithm of frequency instead of the frequency itself. 
If our hypothesis is correct, the rescaled eigenspectra should show a similar pattern, 
just shifted by an amount depending on (the logarithm of)
the individual large separation value. 
These rescaled spectra are presented in figure~\ref{fig-7} where they have 
been shifted according to their individual
$\fmax$ value. Here the sequence of ridges is more visible even if the pattern 
remains blured at low abscissae.
 
\begin{figure*}
\centering
\includegraphics[width=\hsize,clip]{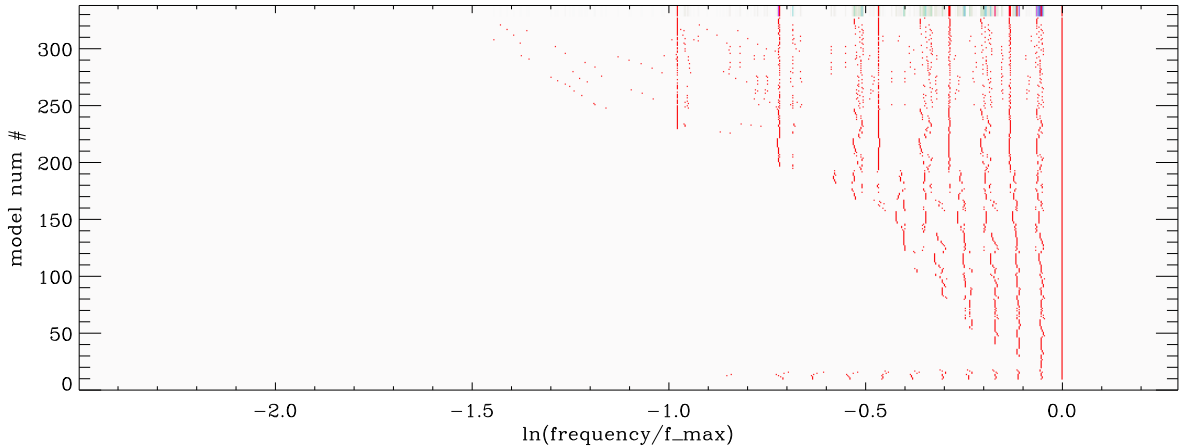}
\caption{
Same as in figure~\ref{fig-7} (except for the 20 bottom raws), but for 
the set of models shown in figure 5, i.e. models
satisfying the same criterion $\fmax > 400\muHz$ than the 
subsample of stars considered in figure~\ref{fig-7}. The 20 bottom raws are
dedicated to \~l=0 axisymmetric modes of models with rotation value going from zero to 
80\% of the breakup rotation rate, as described in Sect.~\ref{sec-31}. 
}
\label{fig-8}       
\end{figure*}

We treated the same way the theoretical spectra associated with
the set of models shown in figure~\ref{fig-5} (models satisfying
the same criterion $\fmax>400\muHz$ than our subsample of stars).
Here again, we considered only unstable axisymmetric (m=0) l=0,1,2 modes.
The result is presented in figure~\ref{fig-8}.

\begin{figure*}
\centering
\includegraphics[width=\hsize,clip]{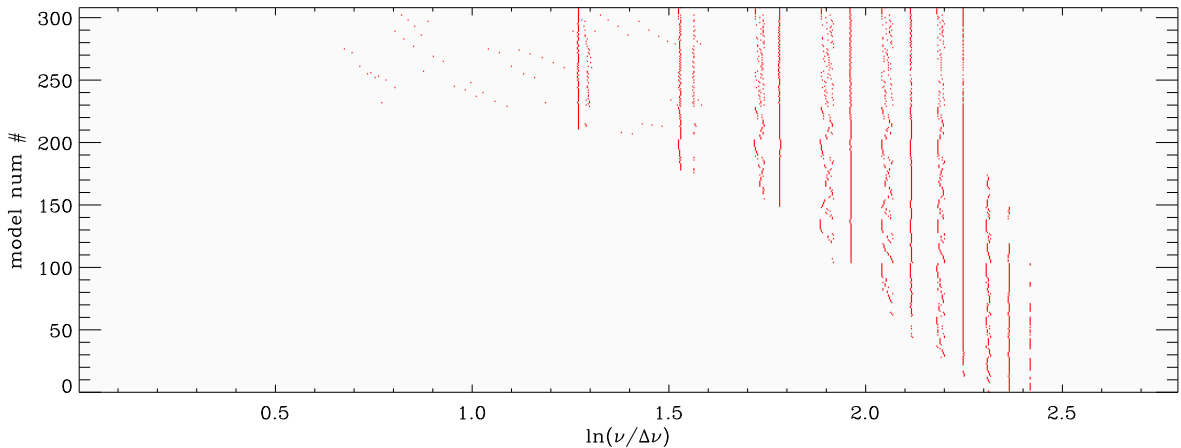}
\caption{
Same as in figure~\ref{fig-8},
but here the
theoretical spectra are normalized by $\dnu$ instead of $\fmax$.
}
\label{fig-9}       
\end{figure*}

The figure~\ref{fig-7} and figure~\ref{fig-8} show great similarities 
with a few clear ridges near zero in abscissae followed by a less clear pattern.
In fact, we have to pay attention to the fact that all these
stars do not have the same range of unstable radial orders and this makes the superposition of 
the spectra less clear. This is demonstrated in figure~\ref{fig-9}, where the 
theoretical spectra this time are normalized by $\dnu$ instead of $\fmax$. 
The ridges are much clearer now. This shows that (for models without rotation) a common
pattern exists and is not too much distorted by the structural changes of models in the early
main sequence.

In order to illustrate the anticipate impact of rotation on this comparison,
we added at the bottom of figure~\ref{fig-8} 
(between line 10 and 18), the pattern 
obtained for axisymmetric (\~l=0) island modes (see \cite{Lignieres2009}) in 9 models based on the 
Self-Consistent Field (SCF) method (\cite{MacGregor2007}).
These models are ZAMS 2$M_{\odot}$, ranging from 0 to 80\% of the break-up rotation rate.
These eigenfrequencies are shown to illustrate the effect of rotation on the pattern, but these
modes are not necessary unstable.   

We see that the regular pattern is preserved to a good extent over this large range of rotation.
In fact, at low radial order, the distribution of the modes appears even more regular than 
the one of models without rotation.

To conclude with this point, it seems important to stress that the spectra considered in this study
obviously show numerous peaks outside of the common quasiregular pattern revealed by figure~\ref{fig-6} or
figure~\ref{fig-7}. On the other hand, these figures do not suggest either that the modes associated to
this pattern are systematically expressed (with detectable amplitude). However these results
suggests that this pattern is characteristic of the eigenspectra of chemically normal 
early main sequence stars.


\section{Conclusions}
\label{sec-con}
We used a homogeneous set of about 1860 stellar light curves collected with CoRoT for
stars classified as 
$\dscu$ stars with a probability higher than 80$\%$ by the automated supervised 
classification of variable stars in the CoRoT archive. 

We have characterized these spectra in terms of range of observed frequencies, defining
two parameters, $\fmin$ and $\fmax$. The distribution of our sample of stars in a
$\fminfmax$ diagram appears to be consistent with the one obtained from a grid of theoretical 
models and linear stability calculations. This suggests that $\fminfmax$ values could
be used to characterize a specific mass range or evolution stage. 

Based on this criterion, we have selected stars that we consider to be on the early main sequence.
We have shown that their spectra reveal a common pattern modulated by individual 
large separation values. The existence of regularities in the spectra of $\dscu$ stars
has already been demonstrated by several studies. It has even been demonstrated that this spacing
is compatible with the classical large separation index used for solar-type pulsators 
and that rotation does not hamper the use of this index.
Here the ridges we find for a large sample of stars confirm these results, but 
in addition, we show on a large set of objects that these 
regularties are due to a
regular pattern of consecutive peaks which is compatible with patterns expected for axisymmetric
island modes as suggested by recent non-perturbative modelling of fast rotating stars 
(\cite{Reese2009}).

We still need to improve our knowledge and parametrisation of this pattern in synergy with further 
theoretical work on fast rotation. 
This study obviously would benefit from being extended to an even larger set of stars.
The extension of this work to $\it Kepler$ data is not expected to change considerably the case
of the early main sequence stage since
the high frequencies characterizing this domain are not accessible with the sampling time
of long cadence data which constitute the essential of $\it kepler$ data. 
However, in the difficult part of the $\fminfmax$ diagram corresponding to evolved $\dscu$ stars,
the $\it Kepler$ data might be of great help.

This suggests that $\fmin$ and $\fmax$ as well as
large separation values 
might be used as seismic indices to characterize
stars and this open the perspective for ensemble seismology using $\dscu$ stars.  


\begin{acknowledgement}
The CoRoT space mission, launched on December 27th 2006,
has been developed and is operated by the Centre National d’Etudes Spatiales (CNES),
with contributions from Austria, Belgium, Brazil, the European Space Agency (RSSD
and Science Programme), Germany and Spain.
We acknowledge the support from the EC Project SpaceInn (FP7-SPACE-2012-312844). 
EM, KB, RS and DR acknowledge the support from the Programme de Physique Stellaire (PNPS).
AGH acknowledges support from Fundação para a Ciência e a Tecnologia (FCT, Portugal) through the
fellowship SFRH/BPD/80619/2011.
JCS acknowledges funding support from Spanish public funds for research under project 
	ESP2015-65712-C5-5-R (MINECO/FEDER), and under Research Fellowship program 
	“Ramón y Cajal” (MINECO/FEDER).
\end{acknowledgement}
 \bibliography{MichelDSC}
%
%
%
%

\end{document}